\documentclass[aps,prl,reprint,superscriptaddress,amsmath,amssymb,floatfix,nofootinbib]{revtex4-2}
\usepackage[T1]{fontenc}
\usepackage[utf8]{inputenc}
\usepackage{graphicx,bm,microtype}
\usepackage[hidelinks]{hyperref}
\usepackage{color}
\usepackage{amssymb}

\hypersetup{pdftitle={Universality of Layered Programmable Decompositions by Fourier Absorption},pdfauthor={Javier Alvarez-Vizoso and David Barral}}

\newcommand{\U}{\mathrm U}

\newcommand{\diag}{\operatorname{diag}}

\newcommand{\tr}{\operatorname{tr}}

\newcommand{\ii}{\mathrm i}

\begin{document}
\title{Universality of Layered Programmable Decompositions by Fourier Absorption}

\author{Javier Álvarez-Vizoso} 
\email{javizoso@gmail.com}
\affiliation{Galicia Supercomputing Center (CESGA), Avda.\ de Vigo S/N, Santiago de Compostela, 15705, Spain}
\author{David Barral} 
\email{david.barral.rana@gmail.com}
\affiliation{Galicia Supercomputing Center (CESGA), Avda.\ de Vigo S/N, Santiago de Compostela, 15705, Spain}
\affiliation{Quantum Materials and Photonics Research Group, Optics Area, \\Department of Applied Physics, iMATUS,
Faculty of Physics, Faculty of Optics and Optometry, \\University of Santiago de Compostela, Santiago de Compostela, 15872, Spain.}

\begin{abstract}

Decomposing arbitrary unitary transformations into physically realizable operations is fundamental to quantum information, quantum control, and linear optics. Layered programmable architectures implement such transformations by alternating tunable phase masks with fixed mode mixers. However, establishing their universality at fixed depth remains challenging: we show that neither a sufficient parameter count nor successful synthesis of sampled targets guarantees global coverage. We prove that an architecture is universal exactly when every reachable transformation can absorb a diagnostic input Fourier multiport by reprogramming its phase masks. The associated Fourier absorption residual (FAR) requires no external target ensemble, vanishes exactly for universal architectures, and bounds worst-target synthesis error. For four-mode Fourier architectures, a permutation obstruction rules out universality with five masks, even under small unitary mixer perturbations, whereas a constructive factorization establishes universality with seven. Beyond exact universality, comparison with a universal reference of the same depth yields certified uniform approximation bounds in terms of unitary mixer errors. Finally, numerical simulations using a geodesic phase optimizer illustrate the distinction between Fourier absorption and direct target synthesis.

\end{abstract}
\maketitle

Unitary transformations are fundamental to mathematics, physics, and engineering, underpinning linear algebra, quantum dynamics, information processing, and communications \cite{CohenI2019,JohnstonII2021,Messiah1961,Arjovsky2016,Zayed2015,Puttnam2021}. In linear optics, they describe both light propagation and changes of mode basis \cite{Miller2019,Fabre2020}. Recent advances in free-space and integrated photonics have enabled precise control of these operations for classical and quantum light \cite{Bogaerts2020, Kim2025}. A widely used approach implements arbitrary unitary transformations through meshes of tunable Mach–Zehnder interferometers (MZIs), each mixing a pair of optical modes using adjustable phase shifters \cite{Reck1994, Clements2016, Bell2021}. These meshes support applications ranging from quantum information processing to optical neural networks \cite{Carolan2015, Bao2023, Maring2024, Zhang2021, Bandy2024, Xie2025}. However, fabrication errors can cause increasing deviations from the target unitary transformation as MZI meshes scale \cite{Pai2019, Burgwal2017,Hamerly2022}. Compensating for these errors can require additional components or more stages of optical processing \cite{Pai2019,Burgwal2017,Hamerly2022,Bandy2021,Miller2013,Miller2017,Mower2015}. Although well suited to spatial modes \cite{Taballione2023, Barzaghi2025}, MZI meshes are more difficult to implement efficiently when information is encoded in frequency or time \cite{Dioum2024, Piao2024}.

\begin{figure*}[t]
\centering\includegraphics[width=0.68\textwidth]{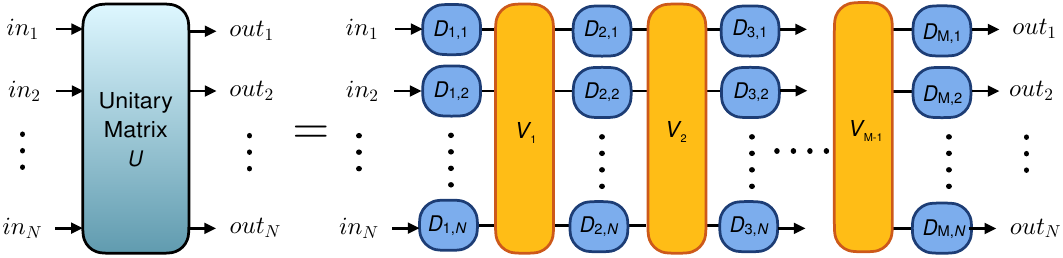}
\caption{Layered programmable decomposition (LPD) of a $N$-dimensional unitary transformation: fixed mixing layers are in yellow and programmable diagonal phase layers are in blue. 
}
\label{fig:car}
\end{figure*}

An alternative is the layered programmable decomposition (LPD), which alternates multimode mixing stages with programmable phase layers (Fig.~\ref{fig:car}) \cite{Morizur2010}. This architecture underlies free-space multi-plane light conversion \cite{Labroille2014, Zhou2018, Fontaine2019, Rouviere2024, Lib2024} and has been adopted in diffractive optical networks \cite{Lin2018, Kulce2021, Ren2026} and integrated optical and microwave processors \cite{Tang2018, Tang2021, Friedman2025, Zelaya2025, Keshavarz2025, Kondratyev2026}. LPD offers a promising route to scalable control through compact implementations, tolerance to loss, and robustness to imperfections \cite{Girouard2025, Tanomura2022, Markowitz2023}. Its compatibility with time- and frequency-domain encoding further extends its reach beyond spatial modes \cite{Ashby2020, Joshi2022, Lukens2017, Budd2021}.

Group-generation results establish that \emph{sufficiently dense} mixing matrices, interleaved with programmable phase layers, can realize any unitary transformation $\mathrm{U}(N)$ through a finite sequence \cite{Borevich1981}. Such results, however, do not establish exact universality at a prescribed depth. Likewise, controllability criteria address reachability with adjustable controls and evolution times \cite{JurdjevicSussmann1972, Dalessandro2009}, while approximate synthesis results such as the Solovay–Kitaev theorem allow variable sequence length \cite{Dawson2005}. Neither directly resolves whether a fixed sequence of mixing stages and phase layers covers the entire unitary group. These questions have motivated extensive theoretical work on which transformations LPD architectures can realize and how to implement them~\cite{Huhtanen2015,Saygin2020,LP2021,Zelaya2024,Yasir2025,Girouard2025,Pereira2025}. Numerical studies have reported high-accuracy synthesis near the parameter-counting threshold of $N+1$ phase masks \cite{Saygin2020,Zelaya2024}, while constructive factorizations for Fourier mixers provide sufficient depths of $2N+5$ and $6N+1$ phase layers for even $N$ \cite{LP2021,Girouard2025}. Beyond these analytical decompositions, numerical and experimental studies typically test whether a finite set of target transformations can be synthesized within a prescribed error tolerance \cite{Morizur2010, Labroille2014, Zhou2018, Fontaine2019, Rouviere2024, Lib2024, Lin2018,Kulce2021, Ren2026, Tang2018, Tang2021, Friedman2025, Zelaya2025, Keshavarz2025, Kondratyev2026, Tanomura2022, Markowitz2023, Ashby2020, Joshi2022, Lukens2017, Budd2021}. Such tests, even when supported by sufficient parameter counts, do not exclude inaccessible transformations, leaving a gap between demonstrated synthesis and guaranteed universality near the minimum depth. The central question is therefore: under what conditions does a prescribed mixer sequence with a fixed number of phase masks realize every unitary transformation?

In this Letter, we establish a global necessary and sufficient criterion based on Fourier absorption for $\mathrm{U}(N)$ universality of LPD at a fixed depth. The criterion asks whether a fixed Fourier transformation appended to every reachable transformation can be absorbed by reprogramming the existing phase layers. 
The resulting Fourier absorption residual (FAR) vanishes exactly for universal architectures and bounds their worst-target synthesis error. For four-mode Fourier architectures, we prove a permutation obstruction with five phase masks that persists under small unitary mixer perturbations, and construct every unitary with seven masks, compared with the thirteen supplied by the general construction of Ref.~\cite{Girouard2025}. Comparison with a universal reference of the same depth bounds the approximation error uniformly over all targets in terms of the unitary mixer errors. We combine the FAR global test with a Riemannian phase optimizer using the exact gradient of the squared unitary geodesic distance. Numerical simulations illustrate absorption and target synthesis as distinct tasks.

\emph{Fixed-depth universality.---}
For $N\ge2$, fix a sequence of static unitary mixing matrices $V_1,\ldots,V_{M-1}$ and let each mask $D_j=\diag(e^{i\phi_{j,1}},\dots,e^{i\phi_{j,N}})$ range independently over the diagonal unitary subgroup $T$ through a set of $N$  programmable phases $\phi_{j, k}$. The LPD architecture (Fig.~\ref{fig:car}) and its reachable image are
\begin{equation}
 L_V({\phi})=D_{M} V_{M-1}\cdots D_{2} V_{1}D_{1},
 \quad S_V=L_V(T^M),
 \label{eq:word}
\end{equation}
with $\phi=(\phi_{1,1}, \dots, \phi_{M,N})$. The question we address here is the global coverage of this image. Universality means $S_V=\mathrm{U}(N)$ with the mixers and depth held fixed. In optical implementations, \(L_V(\phi)\) acts on optical modes, and universality means realizing every unitary mode transformation by tuning the phases.

Removing $M-1$ redundant common phases leaves at most $M(N-1)+1$ effective parameters. Dimensionality therefore requires $M\geq N+1$ \cite{Lee2012}. Although this $M$ provides $N^2$ effective phase parameters, this count does not guarantee universality. For instance, to realize the identity, the final mixer must satisfy $V_N=\tilde{D} C^\dagger$, where $C=D_N V_{N-1}\cdots D_2 V_1$ and $\tilde{D}=D_{N+1}^{\dagger} D_{1}^{\dagger}$. For fixed first $N-1$ mixers, the internal masks supply $(N-1)^2$ relative phases, while $\tilde{D}$ supplies only $N$ effective phases. The compatible final mixers therefore form a smooth image of dimension at most $N^2-N+1<N^2$, and hence a measure-zero subset of $\mathrm{U}(N)$. {Consequently, almost every independently Haar-chosen mixer sequence is nonuniversal at $M=N+1$ since it cannot even realize the identity. This shows why successful numerical target synthesis at the parameter-counting threshold does not establish exact universality \cite{Saygin2020}.} This generic obstruction leaves specially structured sequences, including repeated Fourier mixers, to be assessed separately.

Beyond counting independent parameters, assessing universality at a prescribed depth requires distinguishing local control from global coverage. Full Jacobian rank at a parameter setting guarantees local surjectivity \cite{Lee2012}, while a nonsingular covariance matrix excludes exact affine constraints on the sampled observables \cite{Jolliffe2002}. Neither condition, however, guarantees global coverage of the target unitary group: at a prescribed depth, the reachable set may contain an open neighborhood while still omitting some target transformations. Establishing universality therefore requires an additional global argument, such as formally computing a positive topological degree \cite{outerelo2009mapping}, showing that $S_V$ has empty boundary relative to $\mathrm{U}(N)$ \cite{Munkres2000}, or establishing invariance under a generating set of $\mathrm{U}(N)$. We propose \emph{Fourier absorption}, which implements the latter approach combining the existing diagonal phase freedom with absorption of a discrete Fourier transform, and yields an explicit global residual, amenable to numerical evaluation, whose exact vanishing is equivalent to universality.

\emph{Fourier absorption theorem.---}
Let $(F_N)_{jk}=N^{-1/2}e^{-2\pi i (j-1)(k-1)/N}$, with
$j,k=1,\ldots,N$. For the LPD architecture of Eq. \eqref{eq:word},
\begin{equation}
  S_V=\mathrm{U}(N)
  \quad\Longleftrightarrow\quad
  S_VF_N\subseteq S_V .
  \label{eq:criterion}
\end{equation}
Physically, every reachable transformation preceded by a balanced
Fourier multiport must be reproducible by reprogramming the existing
phase masks, without increasing the depth.

\emph{Proof.---} For sufficiency assume \(S_VF_N\subseteq S_V\): placing a Fourier multiport before any reachable LPD setting can be compensated by reprogramming the existing masks. Besides, an additional input phase mask is automatically absorbed into the input LPD programmable mask (Fig.~\ref{fig:car}), so \(S_V T=S_V\). After each absorption, the reprogrammed LPD is itself at a reachable setting, so the same assumption applies again. Consequently, any finite sequence of phase masks and Fourier transformations can be absorbed into the original LPD
without increasing its depth. Crucially, every unitary transformation can be implemented exactly using a finite sequence of phase masks and Fourier transformations \cite{Morizur2010, Borevich1981}. Choose any reachable transformation $s_0\in S_V$ and any target $U\in\mathrm{U}(N)$. Such a phase mask-Fourier sequence can implement $s_0^\dagger U$, so placing it before the LPD programmed to $s_0$ produces $s_0(s_0^\dagger U)=U$. Thus $S_V=\mathrm{U}(N)$, establishing exact universality at the original depth. For necessity assume \(S_V=\mathrm{U}(N)\). For every reachable setting \(\phi_a\), the transformation \(L_V(\phi_a)F_N\) is unitary. Universality therefore guarantees a setting \(\phi_b\) of the same LPD satisfying \(L_V(\phi_b)=L_V(\phi_a)F_N\). Hence \(S_VF_N\subseteq S_V\). 
$ \blacksquare $

\emph{From absorption to synthesis accuracy.---} We quantify the criterion of universality in Eq. \eqref{eq:criterion} by the \emph{Fourier absorption residual} (FAR)
\begin{equation*}
  \delta_F(V)=
  \max_{\phi_a\in T^M}\min_{\phi_b\in T^M}
  \rho\!\left(L_V(\phi_b),L_V(\phi_a)F_N\right),
  \label{eq:car}
\end{equation*}
with $\rho$ any continuous bi-invariant metric like the normalized chordal distance $\rho_c(A,B)=\|A-B\|_F/\sqrt{2N}$ or the geodesic distance {[Eq. \eqref{eq:geodesicCost}]}. Both extrema are attained by compactness of the phase domain and continuity of the distance. The inner minimization finds the best recovery, while the outer maximization selects the most demanding setting of the same LPD. No external target ensemble is required. Thus,  $\delta_F(V)=0$ is equivalent to exact Fourier absorption for every setting $\phi_a$, with the recovery phases $\phi_b$ allowed to depend on $\phi_a$. 

FAR bounds the worst-target synthesis error, defined as
\begin{equation*}
  E(V)=\max_{U\in\U(N)}\rho(U, S_V).
  \label{eq:worst_error}
\end{equation*}
Suppose that every unitary admits a Fourier LPD factorization using at most $K_N$ Fourier factors \cite{Morizur2010}. For any target $U$, choose a reachable
$s_0\in S_V$ and factor $s_0^\dagger U$. Starting from $s_0$,
diagonal phase masks are absorbed exactly, while each Fourier factor adds at most $\delta_F(V)$ to the recovery error. Bi-invariance and the triangle inequality preserve the accumulated bound, giving $E(V)\le K_N\delta_F(V)$. Conversely, every FAR attempt is a unitary target, so $\delta_F(V)\le E(V)$. Thus, FAR is related to worst-target synthesis error as
\begin{equation}
  \delta_F(V)\le E(V)
  \le K_N \delta_F(V).
  \label{eq:covering}
\end{equation}
For even \(N\), one may take \(K_N=2N+4\) \cite{Girouard2025}. In particular, we prove constructively that every \(U\in\mathrm U(4)\) admits a decomposition with seven phase masks and six identical Fourier $F_4$ mixers (End Matter). This reduces the thirteen masks supplied by the general construction of Ref. [44] to seven and gives \(K_4=6\).  

A certified FAR upper bound $\delta_F(V)\le\varepsilon / K_N$ guarantees synthesis
error at most $\varepsilon$ for every target and, hence, that the architecture is \emph{$\varepsilon$-universal} in the specific metric. Conversely, a
certified lower bound
$\rho(U, S_V)\ge\eta>0$ implies $\delta_F(V)\ge\eta/K_N$. If the witness
is itself a FAR attempt, $U=L_V(\phi_a)F_N$, the stronger bound
$\delta_F(V)\ge\eta$ follows directly, without the Fourier-count
factor. Calibration errors also give a uniform upper certificate. For a universal reference sequence $R$ of the same depth, $E(V)\le\Delta(V,R):=\sum_j\rho(V_j,R_j)$ (End Matter). Thus a device close to a universal design approximates every target with a guaranteed error, without target sampling. For instance, our seven-mask theorem supplies such a reference with six $F_4$ mixers. This calibration error quantifies the robustness of LPD to imperfections \cite{Tanomura2022, Markowitz2023}.

\emph{Riemannian phase optimization.---}
Universality establishes that every unitary target is reachable. Practical synthesis additionally requires finding a corresponding phase configuration. Given $U_{\mathrm{target}}$, we therefore seek the tuple of phases $\phi$ such that $L_V(\phi)=U_{\mathrm{target}}$. A constructive
elimination procedure would directly supply these parameters \cite{Reck1994,Clements2016, Bell2021}.
When no such procedure is available, numerical optimization provides
a practical route.  
Standard optimization methods often minimize a cost function based on the chordal (Euclidean) error
$\rho_c \!\left(L_V(\phi),U_{\mathrm{target}}\right)^2$. A geometrically natural approach is, however, to account explicitly for the curvature of the unitary group by minimizing its intrinsic geodesic
distance \cite{Absil2007}. This distance measures the length of a
shortest path connecting the realized transformation to the target
within $\mathrm{U}(N)$, while the search proceeds through the
programmable phase settings. Riemannian techniques have been used to, e.g., synthesize optimal quantum computing circuits \cite{Nielsen1, Nielsen2}. 

Let $\Omega(\phi)=L_V(\phi)^\dagger U_{\mathrm{target}}$ with orthonormal eigenvectors $v_k$, eigenvalues $\lambda_k=e^{i\theta_k}$, and principal eigenangles $\theta_k\in(-\pi,\pi]$. The cost function can be written as
{\begin{equation}
  f(\phi)=\rho_g \!\left(L_V(\phi),U_{\mathrm{target}}\right)^2
  :=\frac{\sum_{k=1}^{N}\theta_k^2}{2N}
  =\frac{\|\log\Omega(\phi)\|_F^2}{2N} ,
  \label{eq:geodesicCost}
\end{equation}
where $\rho_g$ is the normalized geodesic distance}. Physically, the eigenangles describe the remaining phase errors in the orthogonal eigenmodes of the relative interferometer. Their sum of squares measures the shortest correction within the unitary group and vanishes precisely at an exact fit. Distinctively, we use \emph{eigenvalue derivatives} to guide the phase search, relating each physical phase control to the spectral errors entering the geodesic cost. 
Thus, the exact analytical gradient of $f$ is \cite{Magnus1985}
\begin{equation*}
  \frac{\partial f}{\partial\phi_{j,\alpha}}
  =\frac{1}{N}\sum_{k=1}^{N}\theta_k\,
  \operatorname{Im}\!\left[
    \frac{
      v_k^\dagger
      \left(\partial\Omega/\partial\phi_{j,\alpha}\right)v_k
    }{\lambda_k}
  \right].
  \label{eq:gradGeo}
\end{equation*}
Compared with the squared chordal error, the geodesic objective replaces a $\sin\theta_k$ weighting in the phase gradient by $\theta_k$, avoiding the vanishing spectral weight of errors approaching \(\pm\pi\). In our simulations, we combine multistart Newton–CG optimization with analytical gradients and Hessian–vector products, followed by safeguarded, damped Newton–Raphson refinement using the corresponding analytical Hessian \cite{LPDcode}.

\begin{figure}[t]
\includegraphics[width=0.49\textwidth]{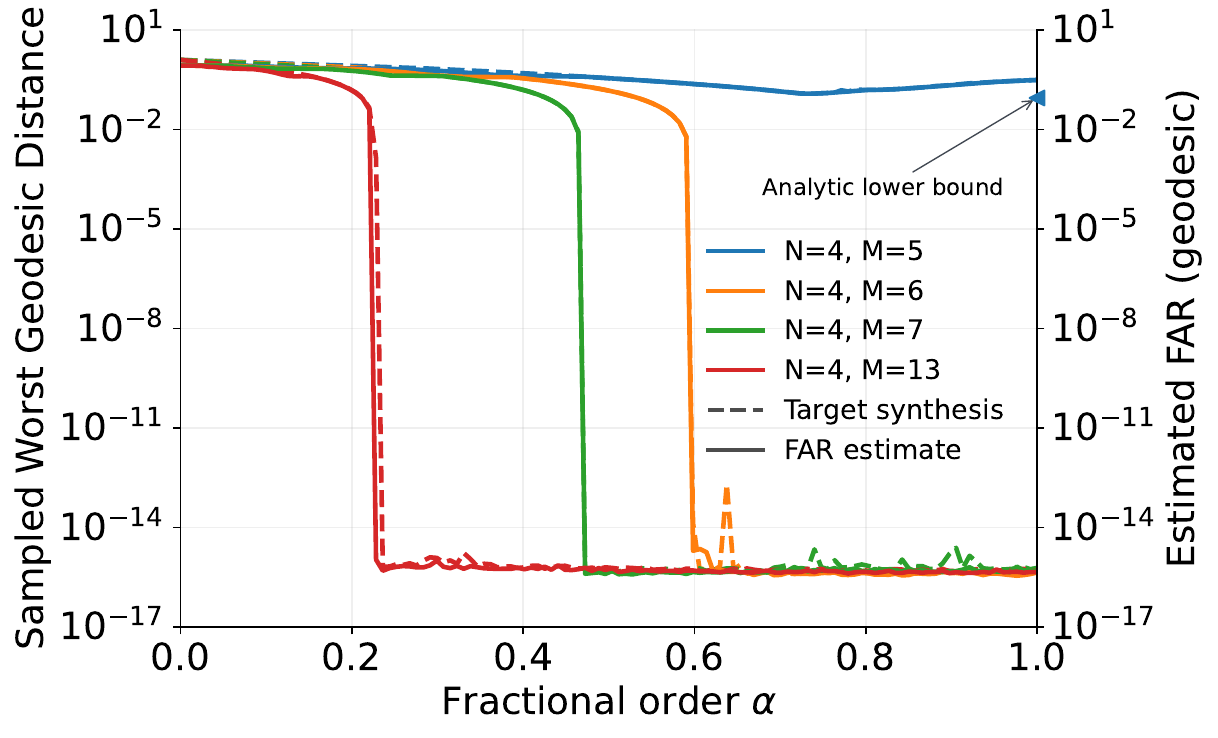}
\caption{Absorption and Riemannian synthesis for fractional-Fourier mixers \(F_N(\alpha)\), with \(N=4\) and \(M=5,6,7,13\). The mixer family interpolates between the identity and the discrete Fourier transform. For each \(\alpha\) and \(M\), the FAR search evaluates 337 random Fourier-displaced reachable targets; solid lines show the largest fitted recovery residual. Dashed lines show the largest synthesis residual among 128 Haar-random, 24 permutation, and 64 block-diagonal targets. FAR tests Fourier-displaced LPD transformations, while synthesis tests external target ensembles. For \(M=6,7,13\), both diagnostics reach numerical precision at similar sampled fractional orders. Fluctuations near numerical resolution are magnified by the logarithmic scale. At \(\alpha=1\) and \(M=5\), an analytical obstruction proves nonuniversality [Eq. \(\eqref{eq:obstruction}\)].}
\label{fig:numerics}
\end{figure}

\emph{Fractional-Fourier simulations.---} Fig.~\ref{fig:numerics} varies all $M-1$ physical mixers along
the fractional DFT family $F_N(\alpha)$, with $F_N(0)=I$ and $F_N(1)=F_N$ \cite{Candan2000}. 
{At small values of $\alpha$, the LPD image is close to a diagonal at every depth limiting reachability and hence universality.} Increasing depth shifts the closely matching drops in the two diagnostics --the worst error of synthesized Monte Carlo sampling and the FAR estimates-- toward smaller fractional orders: {near $\alpha\simeq 0.60,0.47,0.24$ for $M=6,7,13$,} respectively, they both drop to machine precision. For $M=5$, both diagnostics remain positive throughout the sampled
sweep, in accordance with the analytical obstruction that proves nonuniversality at the Fourier endpoint $\alpha=1$ (Eq. \eqref{eq:obstruction}, Fig.~\ref{fig:numerics} inset). Solid curves test Fourier-displaced reachable matrices $L_V(\phi_a)F_4$, while dashed curves show the largest error recovered among Haar, permutation, and block-diagonal targets. Both diagnostics use the normalized geodesic distance $\rho_g$. The matching transition regions show that Fourier absorption tracks the onset of high-accuracy synthesis across structurally distinct target ensembles. {Furthermore, for $\alpha=1$ at $M=7$ and $13$, and $\alpha=0.5$ at $M=13$ \cite{PER3}, machine-precision FAR and Riemannian synthesis align with exact universality \cite{Girouard2025}.} 

These numerical experiments illustrate FAR as an architecture diagnostic that does not require selecting an external target ensemble. However, they are not certified bounds: incomplete outer sampling can underestimate FAR, while imperfect inner minimization can overestimate it. Independent uniform phase sampling can additionally establish $\epsilon$-approximate Fourier absorption over a fraction of phase volume with a given statistical confidence (End Matter). Below, we establish explicit analytical certificates of nonuniversality and prove their persistence under small mixer perturbations with an example.

\emph{An exact obstruction at minimal depth.---} A positive FAR lower bound suffices to certify nonuniversality. Consider $N=4$, $M=5$, and four Fourier mixers $V_j=F:=F_4$.
Although $M(N-1)+1=16=\dim\U(4)$, this architecture misses an
explicit family of permutations. The demonstration is as follows: for each diagonal mask $D$, the unitary $FDF^\dagger$ is
circulant, which means it commutes with the cyclic shift
$Xe_j=e_{j+1\bmod4}$. We use that $FDF=(FDF^\dagger)J$ with $J=F^2$, and the fact that conjugation by $J$ preserves diagonal and circulant matrices. Hence, every reachable output has the form $U=D_LWD_R$, with $W=C_L D C_R$ and where $C_L, C_R$ are circulant unitaries, and $D_L,D,D_R$ are
diagonal unitaries. Commutation with $X$ and cyclicity of the trace thus yield the following relation
\begin{equation}
 g(W):=\tr(WXW^\dagger X)
      =\tr(DXD^\dagger X)=0.
 \label{eq:four-trace}
\end{equation}
Every input mode is shifted by two positions. No basis vector returns to itself, so every diagonal entry vanishes.

Now, let us define a permutation matrix $P$ that exchanges modes $3$ and $4$, fixing $1$ and $2$. Removing the endpoint masks gives
$Q=D_L^\dagger P D_R^\dagger=\diag(q_1,q_2,q_3,q_4)P$,
with $|q_j|=1$.
Direct evaluation yields $g(Q)=q_3\overline q_4$, and hence
$|g(Q)|=1$. Therefore, the contradiction with Eq. \eqref{eq:four-trace} implies that $P$ cannot be reachable and that this specific LPD architecture is nonuniversal.

{This conclusion is also captured by our method.} Expanding $g(Q)-g(W)$ and applying the Frobenius
Cauchy--Schwarz inequality which gives
$|g(Q)-g(W)|\le4\|Q-W\|_F =4\|P-U\|_F$, we obtain $\rho_c(P, S_{F_4})\ge\eta:=1/(8\sqrt2)$. Then \cite{PER}
\begin{equation}
  E(F_4)\ge \delta_F(F_4) \ge \eta=1/(8\sqrt2).
 \label{eq:obstruction}
\end{equation}
This obstruction yields a quantitative ceiling on the process fidelity $\mathcal F$ for implementing $P$: \(\rho_c(P,S_{F_4})^2=1-\sqrt{\mathcal F_{\max}(P)}\), and hence \(\mathcal F_{\max}(P)\le(127/128)^2\simeq98.44\%\).
Eq. \eqref{eq:obstruction} holds for both normalized chordal and geodesic distance \cite{PER2}, and proves nonuniversality independently of numerical optimization. Together with the seven-mask construction, the obstruction bounds the minimum number of phase masks for four-mode Fourier architectures as \(6\le M_{\min}\le7\). Universality with six masks remains an open question \cite{Pereira2025}.

{
The obstruction of Eq. \eqref{eq:obstruction} persists under unitary mixer perturbations.
For a sequence of mixers $\widetilde V=(\widetilde V_1,\ldots,\widetilde V_4)$,
we define $\Delta=\sum_{j=1}^4\|\widetilde V_j-F_4\|_F/\sqrt8$.
Telescoping at identical phase settings gives
$\rho_c(L_{\widetilde V}(\phi),L_F(\phi))\le\Delta$.
The triangle inequality preserves the fixed-target gap up to
$\Delta$ and the FAR gap up to $2\Delta$. The seven mask theorem gives $E(\widetilde V)\le 6 \delta_F (\widetilde V)$ for arbitrary perturbed mixers, yielding (End Matter)
\begin{equation}
\begin{aligned}
E(\widetilde V)
&\ge \rho_c(P,S_{\widetilde V})\ge[\eta-\Delta]_+,\\
\delta_F(\widetilde V)
&\ge\max\left\{[\eta-2\Delta]_+,
\frac{[\eta-\Delta]_+}{6}\right\},
\end{aligned}
\label{eq:robust-obstruction}
\end{equation}
with $[x]_+=\max\{x,0\}$. Hence, $\sum_j\|\widetilde V_j-F_4\|_F<1/4$ defines an open neighborhood of nonuniversal architectures. These bounds rule out exact universality at the parameter-counting threshold, previously conjectured on the basis of high-fidelity numerical synthesis for a limited set of targets \cite{Saygin2020,Zelaya2024}.
}

\emph{Conclusion.---}
We have established a global characterization of universality for LPD with arbitrary ordered mixers at fixed depth. Universality is equivalent to the ability to absorb a diagnostic Fourier transformation applied to every reachable operation by reprogramming the existing phase masks. The resulting Fourier absorption residual vanishes precisely when the reachable image is the full unitary group, providing a necessary and sufficient criterion that does not rely on an external ensemble of target unitaries.
The accompanying Riemannian optimizer addresses the complementary task of finding phase settings for a prescribed architecture and target transformation, while preserving the geometry of the unitary group.  
The four-mode obstruction shows that parameter count and Fourier mixing alone do not ensure global coverage, even in a neighborhood of
the nominal design. {The complementary seven-mask construction bounds the remaining depth question to six or seven masks.}

Our criterion applies to programmable mode transformations in spatial, temporal, and frequency platforms \cite{Tang2018,Tang2021,Friedman2025,Zelaya2025,Keshavarz2025,Kondratyev2026,Ashby2020,Joshi2022,Lukens2017,Budd2021}. It can assess whether a prescribed device covers the transformations required for mode sorting, communications, diffractive neural networks or quantum information protocols \cite{Labroille2014, Zhou2018, Fontaine2019, Lin2018, Kulce2021, Ren2026, Lib2024, Rouviere2024}. Complete control of optical mode transformations is relevant to non-Gaussian entanglement detection \cite{Zhang2023, Tian2025, Barral2024, Lopetegui2025} and entanglement routing \cite{Fainsin2025}. Assessing the optical architectures used in these settings, and extending the approach to other factorizations \cite{Peruzzo2025, Markowitz2025}, are potential directions for future work. {The analytical geodesic-error gradient provides a phase-synthesis tool with explicit derivatives for programmable optical architectures}, including multi-plane light conversion \cite{Philips2023, Sureka2026} and optimization of heralded state preparation \cite{Parellada2026}. The calibration certificate connects exact constructions to uniform synthesis guarantees for
imperfect devices. Further work should determine minimal universal depths beyond four modes and develop
structural certificates where no nearby universal reference is known. Together, those studies would complement the present toolbox and support the robust design of universal LPD interferometers.

\emph{Acknowledgements.} The authors thank A. Gómez and D. García-Selfa for insightful discussions during the early stages of this project, and M. Walschaers and two anonymous reviewers for valuable comments on previous versions of the manuscript. J.A.-V. was supported by the Ministry for Digital Transformation and of Civil Service of the Spanish Government through the QUANTUM ENIA project call---Quantum Spain project, and by the European Union through the Recovery, Transformation and Resilience Plan---NextGenerationEU within the framework of the ``Digital Spain 2026 Agenda''. D.B. was supported by MICINN through the European Union NextGenerationEU recovery plan (PRTR-C17.I1) and the Galician Regional Government through ``Planes Complementarios de I+D+I con las Comunidades Autónomas'' in Quantum Communication, and by grant RYC2024-048797-I funded by MICIU/AEI/10.13039/501100011033 and by ESF+.

\emph{Data availability.} The data that support the findings of this article are openly available in \cite{LPDcode}.


\clearpage

\appendix
\begin{center}\textbf{End Matter}\end{center}
\label{sec:endmatter}

\section*{Seven-mask universality for four modes}

In the following, we prove that any four-mode unitary can be implemented with seven
phase masks and six identical Fourier mixers. 

\emph{Cosine--sine factorization.---} 
Block matrices are written in the order $(1,3\mid2,4)$. The cosine--sine decomposition writes any unitary target as~\cite{Sutton2009}
\begin{equation}
 U=K_L R(\alpha,\beta)K_R,\qquad
 R=\begin{pmatrix}c&-s\\s&c\end{pmatrix},
 \label{eq:csd}
\end{equation}
where $K_L,K_R$ act independently within the two pairs,
$c=\diag(\cos\alpha,\cos\beta)$,
$s=\diag(\sin\alpha,\sin\beta)$, and
$\alpha,\beta\in[0,\pi/2]$. Thus $R$ consists of two independent
beam splitters connecting the pairs. These two mixing angles can
be supplied by the circulant
\begin{equation*}
 C(\alpha,\beta)=F\diag(e^{\ii\alpha},e^{\ii\beta},
                           e^{-\ii\alpha},e^{-\ii\beta})F^\dagger.
 \label{eq:central}
\end{equation*}
Expressing the circulant in this block basis gives $C(\alpha,\beta)=QRQ^\dagger$,
with the fixed pair transformation
\begin{equation*}
 Q=\frac{1}{\sqrt2}
 \begin{pmatrix}
  1&1&0&0\\
  1&-1&0&0\\
  0&0&\ii&1\\
  0&0&\ii&-1
 \end{pmatrix}.
 \label{eq:basis}
\end{equation*}
Absorbing $Q$ into the pair transformations in Eq.~\eqref{eq:csd}
therefore gives 
\begin{equation*}
U=K'_L C(\alpha,\beta)K'_R.
\end{equation*}

\emph{Common pair mixers.---} 
We next show that both $K'$ transformations can be written in terms of circulant and diagonal matrices.
Each factor $K'$ contains two independent $U(2)$ transformations $K'=W_1\oplus W_2$, acting on the pairs \(\{1,3\}\) and \(\{2,4\}\).  

A lossless two-mode mixer can be written as 
\[
 x(a)=e^{\ii a\sigma_x}
 =\begin{pmatrix}
 \cos a&\ii\sin a\\
 \ii\sin a&\cos a
 \end{pmatrix}.
\]
We apply this rotation to each $W_{j=1,2}$ as $Y_j=x(-a)W_j$. The output power at the first output of each block is
$p_j(a)=|(Y_j)_{11}|^2$.
Increasing $a$ by $\pi/2$ exchanges the two output amplitudes up to
phases, so 
\begin{equation*}
 p_j(a+\pi/2)=1-p_j(a).
 \label{eq:balance}
\end{equation*}
The continuous function $f=p_1-p_2$ satisfies $f(a+\pi/2)=-f(a)$,
so it has a zero in between. Thus there is a common $a$ such that $p_1(a)=p_2(a)\equiv p$.
The two transformed blocks then have the same power splitting
ratio $p$ and unitarity fixes the other three entry magnitudes
\[
 \begin{pmatrix}p&1-p\\1-p&p\end{pmatrix}.
\]
Thus both transformed devices have the same power-transfer probabilities.
Their amplitudes may still have different phases, but those differences
can be removed with input and output phase shifts. $U(2)$ unitarity implies 
\[
Y_j=d_j\,x(b)\,e_j,
\qquad \cos^2 b=p,
\]
where \(d_j,e_j\) are diagonal unitaries. The angle $b$ is the same
for both devices, whereas their phase settings can differ. Since $Y_j=x(-a)W_j$, multiplying on the left by $x(a)$ gives
\begin{equation}
 W_j=x(a)d_jx(b)e_j,
 \label{eq:paired}
\end{equation}
where $a,b$ are common to both
pairs. The common pair mixers follow a circulant identity in physical mode order $(1,2,3,4)$ \cite{Girouard2025}
\begin{gather*}
 x(a)\oplus x(a) = F\diag(e^{\ii a},e^{-\ii a},e^{\ii a},e^{-\ii a})F^\dagger \equiv T(a), \\
  x(b)\oplus x(b) = F\diag(e^{\ii b},e^{-\ii b},e^{\ii b},e^{-\ii b})F^\dagger \equiv T(b).
 \label{eq:paired-circulant}
\end{gather*}
Combining the two blocks in Eq.~\eqref{eq:paired} therefore gives
\[K'=T(a)D T(b)E,\]
with $D,E$ diagonal $U(4)$ unitaries. Taking adjoints also provides the reversed form $K'=D T(a)E T(b)$,
with relabeled angles and phases.

\emph{From circulant to Fourier LPD.---}
Finally, applying these factorizations to $K'_L$ and $K'_R$ we get
\begin{equation*}
 U=A_4 T(a_L)A_3
 \underbrace{[T(b_L)C(\alpha,\beta)T(a_R)]}_{C_2}
 A_2 T(b_R)A_1,
 \label{eq:merge}
\end{equation*}
where $A_1,\ldots,A_4$ are diagonal unitaries.
The bracket is a single unitary circulant because all its factors
are diagonal in the Fourier basis. With $C_3=T(a_L)$ and $C_1=T(b_R)$,
this proves the exact factorization
\begin{equation}
                    U=A_4C_3A_3C_2A_2C_1A_1.
 \label{eq:seven}
\end{equation}
Then we substitute $C_j=F\Lambda_jF^\dagger$, with $\Lambda_j$
diagonal. Since $J=F^2$ obeys $J^2=I$, $F^\dagger=FJ$, and $JDJ$
is diagonal, all three $J$ factors can be moved to the right,
leaving one final $J$. Applying Eq.~\eqref{eq:seven} to the target
$UJ$ cancels this reversal and yields
\begin{equation*}
              U=D_7FD_6FD_5FD_4FD_3FD_2FD_1.
\end{equation*}
Thus, seven phase masks suffice to synthesize every $U\in U(4)$ and their phases can be obtained analytically.

\section*{Statistical validation of approximate Fourier absorption}
\label{em:stat-section}

For a fixed architecture and prescribed tolerance $\epsilon$, we set 
$n$ phase configurations $\phi_a$ independently and uniformly. If configurations whose optimal recovery residual
\[ r(\phi_a)=\min_{\phi_b}\rho\!\left(L_V(\phi_b),L_V(\phi_a)F_N\right) \] 
exceeds \(\epsilon\) occupy a normalized phase volume of at least $q$, the probability of missing this failing set in all $n$ tests satisfies $P_{\rm miss}\le(1-q)^n$. Setting e.g. $q=10^{-4}$ and requiring $P_{\rm miss}<10^{-4}$, gives $n\ge92\,099$. An all-success run with valid numerical recoveries therefore supports absorption within $\epsilon$ over at least $99.99\%$ of the phase parameter volume, with $99.99\%$ confidence. As an example, a LPD with fractional Fourier mixers, $N=4$, $M=6$, and  $\alpha=0.7$, passed all \(92\,099\) independent tests for $\epsilon = 10^{-12}$ \cite{LPDcode}. 

\section*{Certified absorption and calibration bounds}
\label{em:robust-section}

For an LPD image $S_V \equiv S$ [Eq. \eqref{eq:word}], we define 
\begin{equation*}
 \begin{aligned}
 &\rho(U,S)=\min_{s\in S}\rho(U,s),\\
 &E(S)=\max_U\rho(U,S), \\
 &\delta_F(S)=\max_{s\in S}\rho(sF,S),
 \end{aligned}
\end{equation*}
with $\rho$ any continuous bi-invariant metric. The image is compact and satisfies $ST=S$.

\emph{Completion at fixed depth.---}
We write $S^{[k]}=S(FT)^k$. Starting from any $s\in S$, we absorb the additional Fourier factors one at a time, choosing a nearest recovery in $S$ after each factor. Diagonal masks are absorbed exactly. Bi-invariance and the triangle inequality bound the accumulated error
by $k\delta_F(S)$. Hence, every $x\in S^{[k]}$ satisfies
$\rho(x,S)\le k\delta_F(S)$, and
\begin{equation}
 \begin{aligned}
 \rho(U,S)&\le\rho(U,S^{[k]})+k\delta_F(S),\\
 E(S)&\le E(S^{[k]})+k\delta_F(S).
 \end{aligned}
 \label{em:completion}
\end{equation}

{\emph{Calibration transport.---}
For equal-length mixer sequences $V,R$, let
$\Delta(V,R)=\sum_j\rho(V_j,R_j)$.
Replacing the mixers one at a time at identical phase mask settings we obtain
}\begin{equation*}
 \rho(L_V(\phi),L_R(\phi))\le\Delta(V,R)
 \quad\text{for every }\phi.
 \label{em:robust-telescope}
\end{equation*}
Each replacement contributes exactly its mixer distance because the
surrounding factors are unitary. Using the same phase setting in the two devices and minimizing, we get $|\rho(U,S_V)-\rho(U,S_R)|\le\Delta(V,R)$.
Then, maximizing over $U$ gives the first bound below. For FAR, both the attempt $L_V(\phi_a)F$ and the recovery $L_V(\phi_b)$ move by at most $\Delta(V,R)$. Thus, taking the inner minimum and outer maximum we attain
\begin{equation}
 \begin{aligned}
 |E(V)-E(R)|&\le\Delta(V,R),\\
 |\delta_F(V)-\delta_F(R)|&\le2\Delta(V,R).
 \end{aligned}
 \label{em:calibration-stability}
\end{equation}
If $R$ is universal, $E(R)=0$ proves
$E(V)\le\Delta(V,R)$. Equivalently, the same masks that exactly
implement a target in $R$ approximate it within $\Delta(V,R)$ in $V$.

{\emph{Persistence of the four-mode obstruction.---}
Now, we take four nominal mixers $F_4$ and perturbed unitary mixers
$\widetilde V_j$, with the chordal budget
}\[
 \Delta=\frac1{\sqrt8}\sum_{j=1}^4\|\widetilde V_j-F_4\|_F.
\]
The nominal worst-target error and FAR gaps are by Eq. \eqref{eq:obstruction} at least $\eta=1/(8\sqrt2)$.
Equation~\eqref{em:calibration-stability} and its targetwise version
therefore give
\begin{equation}
 \begin{aligned}
 E(\widetilde V)&\ge\rho_c(P,S_{\widetilde V})
                    \ge[\eta-\Delta]_+,\\
 \delta_{F}(\widetilde V)&\ge[\eta-2\Delta]_+,
 \end{aligned}
 \label{em:robust-synthesis}
\end{equation}
where $P$ is the permutation that exchanges modes 3 and 4. In particular, for $\Delta<\eta$ we get
\begin{equation*}
 \sum_{j=1}^4\|\widetilde V_j-F\|_F<\frac14,
 \label{em:robust-neighborhood}
\end{equation*}
which defines an open neighborhood of nonuniversal architectures.
For $\Delta\ge\eta$, this estimate gives no conclusion about
universality.

Finally, for any $s_0\in S_{\widetilde V}$ and any target $U$, the seven-mask
theorem factors $s_0^\dagger U$ with six Fourier {mixers. Its leftmost
mask is absorbed into $s_0$, giving
$S_{\widetilde V}(FT)^6=\mathrm U(4)$.
Equation~}\eqref{em:completion} {thus implies
$E(\widetilde V)\le6\delta_{F}(\widetilde V)$ and
}\begin{equation}
 \delta_{F}(\widetilde V)\ge
 \max\left\{[\eta-2\Delta]_+,\frac{[\eta-\Delta]_+}{6}\right\}.
 \label{em:robust-combined}
\end{equation}
The first term is stronger near zero perturbation while the second
remains positive throughout $\Delta<\eta$.
Finally, $\rho_g\ge\rho_c$ \cite{PER2} transfers the synthesis and FAR lower bounds to geodesic metric, with the same chordal perturbation budget $\Delta$.


\begin{thebibliography}{99}
\bibitem{PER3} Two half-order fractional Fourier mixers separated by an identity mask implement the DFT up to an absorbable scalar phase. The thirteen-mask half-order architecture therefore contains the universal seven-mask Fourier architecture as the sub-family obtained by fixing alternate internal masks to the identity.
\bibitem{PER} {Taking $A=\operatorname{diag}(1,e^{-i\frac{\pi}{4}},i,e^{i\frac{\pi}{4}})$, $B=\operatorname{diag}(1,-i,1,-i)$, and $R=\operatorname{diag}(1,e^{3i\frac{\pi}{4}},1,e^{-3i\frac{\pi}{4}})$; we have $P = s F R$ with $s=AFBFA^\dagger FAFB^\dagger\in S_{F}$. Since $S_{F}R=S_{F}$, bi-invariance gives
$\rho(sF_4,S_{F})=\rho(P,S_{F})$.}
\bibitem{PER2} {$\rho_c(A,B)^2=\sum_k 4\sin^2 (\theta_k /2)/(2N) \le \sum_k \theta_k^2 /(2N) =\rho_g(A,B)^2$, where $e^{i\theta_k}$ are the eigenvalues of $A^\dagger B$ and max-min preserves this ordering.}
\end{thebibliography}
\end{document}